\documentclass[pra,showpacs,preprint,superscriptaddress,amsmath,amssymb]{revtex4-1}

\setcitestyle{round}

\usepackage{graphicx}
\usepackage{dcolumn}

\usepackage[usenames]{color}

\begin{document}
\raggedbottom

\title{Non-adiabatic quantum interference effects and chaoticity in the ultracold Li + LiNa $\to$ Li$_2$ + Na reaction}
\author{Brian K. Kendrick}
\email{Correspondence should be addressed to BKK (bkendric@lanl.gov).}
\affiliation{Theoretical Division (T-1, MS B221), Los Alamos National Laboratory, Los Alamos, NM 87545, USA}
\author{Hui Li}
\author{Ming Li}
\author{Svetlana Kotochigova}
\affiliation{Department of Physics, Temple University, Philadelphia, PA 19122, USA}
\author{James F. E. Croft}
\affiliation{Department of Physics, The Dodd-Walls Centre for Photonic and Quantum Technologies,  University of Otago, Dunedin, New Zealand}
\author{Naduvalath Balakrishnan}
\affiliation{Department of Chemistry and Biochemistry, University of Nevada, Las Vegas, NV 89154, USA}
\vskip 10pt
\begin{abstract}

Electronically non-adiabatic effects play an important  role in many chemical reactions. How  these effects manifest in cold and ultracold chemistry remain largely unexplored. Here, through first principles non-adiabatic quantum dynamics calculations of the Li + LiNa $\to$ Li$_2$ + Na chemical reaction, it is shown that  non-adiabatic dynamics induces quantum interference effects that dramatically alter the ultracold rotationally resolved reaction rate coefficients. The interference effect arises from a conical intersection between the ground and an excited electronic state that is energetically accessible even for ultracold collisions. These unique interference effects might be exploited for quantum control applications as a quantum molecular switch.  A statistical analysis of rotational populations of the Li$_2$ product reveals a Poisson distribution implying an underlying classically chaotic dynamics.
The Poisson distribution is robust and amenable to experimental verification and appears to be a universal property of ultracold reactions involving alkali metal dimers.
\end{abstract}

\maketitle

{\color{red} \section*{Introduction}}

Ultracold molecules and in particular ultracold polar molecules are at the forefront of precision spectroscopy, sensing, controlled studies of chemical reactions,
quantum many-body physics, and quantum computing\cite{Carr_Review2009,OspelkausExpKKRb2010,bala2018,PerreaultScience2017,Rvachov2017,Rui_ExpNaK2017,PerreaultNatChem2018,croftbalahua2018,Ye_ExpNaRb2018,Hu_KRb2019}.
Polar molecules comprised of heteronuclear alkali metal dimers such as KRb, NaK, NaRb and LiNa have attracted considerable attention in recent years in controlled studies of chemical reactions\cite{OspelkausExpKKRb2010,Rvachov2017,Rui_ExpNaK2017,Ye_ExpNaRb2018,Hu_KRb2019}.
Electronically non-adiabatic effects are expected to play an important role in atom-dimer reactions involving these molecules.
The reactions proceed along a barrierless reaction pathway into a deep attractive potential well.
A conical intersection (CI) occurs between the ground electronic state and the first excited doublet electronic state within the attractive well region
and this CI is energetically accessible even for collision energies in the utracold limit for ground state reactants.
Thus, a non-adiabatic quantum mechanical treatment is required that includes both electronic states.
Explicit quantum calculations for these reactions remain a formidable challenge even for dynamics on a single Born-Oppenheimer adiabatic electronic potential energy
surface (PES)\cite{makrides2015,croftNatCom2017,croftPhysRev2017}.
Fortunately, we have recently developed a new quantum reactive scattering methodology that has made it possible to treat non-adiabatic ultracold reactions occurring on
two coupled electronic states for the first time\cite{Kendrick2018nonad}.

In this work, we present a first principles full-dimensional quantum dynamics study of non-adiabatic effects in the
Li + LiNa($v=0$, $j=0$) $\to$ Li$_2$($v'$, $j'$) + Na reaction.
The rotationally resolved rate coefficients are computed as a function of collision energy from $1\,{\rm nK}$ to $10\,{\rm K}$ using a coupled two-state diabatic electronic representation\cite{Kendrick2018nonad,Kendrick2018nonadHH2,Kendrick2019nonadHHD}.
The non-adiabatic results are compared to a conventional Born-Oppenheimer calculation based on a single adiabatic electronic PES.
Both of these calculations are also compared to a universal model which is based on a simple one-dimensional reaction path consisting of a long-range van der Waals (C$_6$) potential\cite{C6_Hui2019}.
Quantum interference between the two reaction pathways which encircle the CI is shown to significantly enhance or suppress the rate coefficients at ultracold collision energies (i.e., $E_c < 1\,{\rm mK}$).
The geometric phase (GP) which is included in the non-adiabatic calculations reverses the nature of the quantum interference from constructive to destructive and vice
versa\cite{natcom2015,PRL2015,Hazra2015HO2}.
Thus, the non-adiabatic ultracold rate coefficients are significantly enhanced or suppressed relative to the conventional Born-Oppenheimer rates coefficients when quantum interference effects are significant.
The quantum dynamics calculations are based on accurate {\it ab initio} electronic PESs which are computed for both the ground and first excited states for the first time.
A state-of-the-art electronic structure code (MOLPRO) is used to compute the electronic PESs and the non-adiabatic coupling elements\cite{molpro}.
Strong fluctuations are observed in the rotationally resolved rate coefficient distributions.
A statistical analysis of these fluctuations reveals that they are Poissonian which is consistent with an underlying classically chaotic dynamics\cite{croftNatCom2017,croftPhysRev2017}.
The Poisson distributions are shown to be robust with respect to variations in the PES and chemical system and therefore appear to be a universal property of these types of reactions that proceed  through  a potential well.

{\color{red} \section*{Results}}
\subsection*{Potential Energy Surfaces for Li$_2$Na}
\begin{figure}[h]
\includegraphics[scale=2.0]{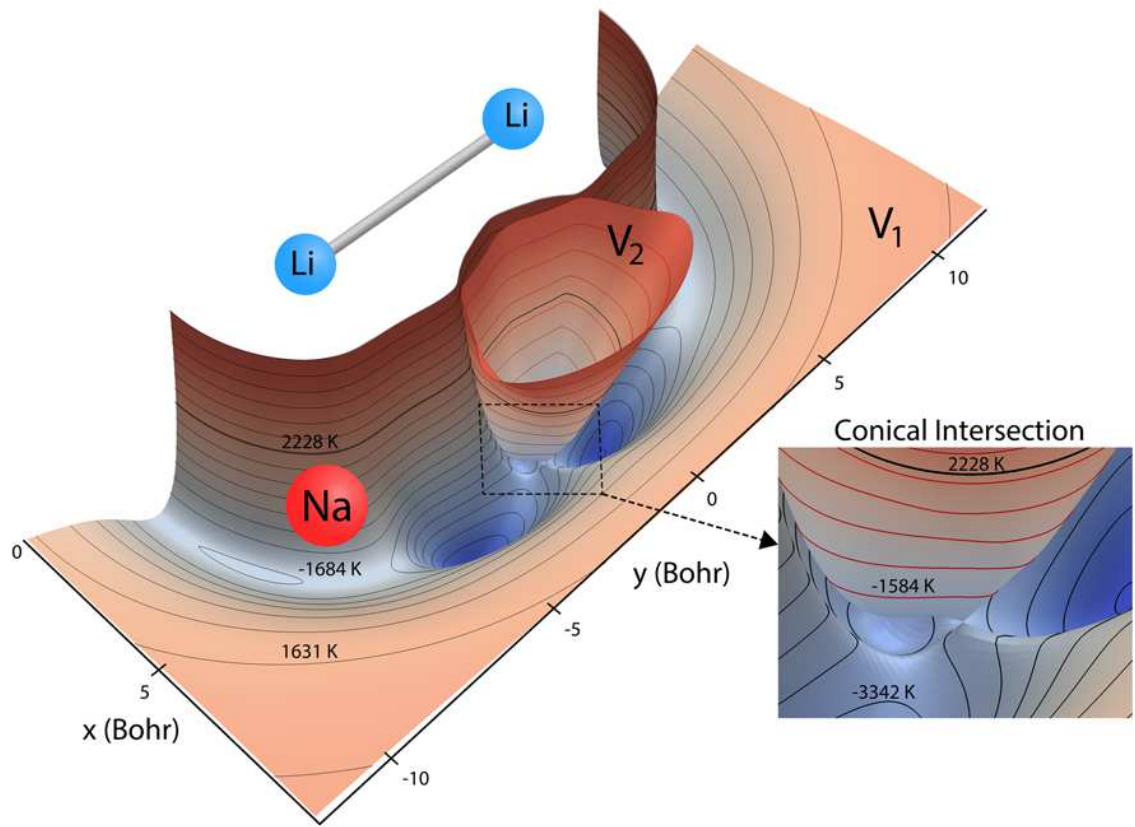}
\caption{
  \begin{small} { \textbf{Potential energy surfaces of Li$_2$Na.}
{\it Ab initio} Born-Oppenheimer potential energy surfaces (PESs) for Li$_2$Na are plotted for a fixed Li$_2$ bond length of $6.25\,{\rm Bohr}$. The $xy$ coordinates denote the location of the sodium nuclei (red sphere) relative to the center of the bond between the two Li nuclei (blue spheres). The ground ($V_1$) and excited ($V_2$) electronic state surfaces are contoured with black and red contours, respectively. Two attractive potential well regions (blue) are clearly visible on the ground state surface. The excited state surface exhibits a conical intersection with the ground state surface for T-shaped ($C_{2v}$) geometries (see inset). The thick black contour line denotes the total energy $2\,228\,{\rm K}$ of the reactant Li + LiNa($v=0$, $j=0$). The other $20$ black contours lie between $-5\,000\, {\rm K}$ and $5\,500\,{\rm K}$ inclusive. The $10$ red contour lines lie between $-2\,470\,{\rm K}$ and $5\,500\,{\rm K}$ inclusive. Both surfaces are not plotted above $5\,500\,{\rm K}$. 
}
\end{small}
}
\label{fig01}
\end{figure}

The Born-Oppenheimer electronic PESs are plotted in Fig.~\ref{fig01} for both the ground and first excited electronic states of the Li$_2$Na molecule.
These surfaces are computed in full dimensionality (i.e., as a function of all three bond lengths) from first principles (see {\bf Materials and Methods} for details)\cite{molpro}.
The PESs in Fig.~\ref{fig01} are two-dimensional slices plotted for a fixed Li$_2$ bond length of $6.25\,{\rm a}_{\rm 0}$ (close to its equilibrium bond length)
and show the topology of the effective interaction potential experienced by the Na nuclei in the vicinity of Li$_2$.
Notable features include the two deep attractive wells (blue colored regions) on the ground state surface (black contours) and the inverted cone of the excited electronic state (red contours).
All energies are reported relative to the bottom of the asymptotic potential well for the Li$_2$ + Na product channel.
The minimum energy of the symmetric potential wells is $-5\,814\,{\rm K}$
(see the thick solid black curve in Fig.~S1 in {\bf Supplementary Materials}).
The ground and excited state PESs exhibit a conical intersection for T-shaped (i.e., $C_{2v}$) geometries (see Fig.~\ref{fig01} inset).
The minimum energy of the conical intersection is $-3\,140\,{\rm K}$ (see the thick solid red curve in Fig.~S1).
The asymptotic energy of the Li +  LiNa($v=0$, $j=0$) reactant channel is shown by the thick black contour line at $2\,228\,{\rm K}$ (see also the thick horizontal dashed line in Fig.~S1).
From Fig.~\ref{fig01} we see that for ultracold collisions of Li with LiNa in its ground vibrational and rotational state, {\it both} the ground and excited electronic states are energetically accessible in the interaction region.
Thus, both electronic states and the couplings between them must be included in the quantum dynamics calculations (see {\bf Materials and Methods} for details)\cite{Kendrick2018nonad}.
These couplings include the GP associated with the conical intersection shown in Fig.~\ref{fig01}.
As discussed in detail in the following section, the GP can lead to a dramatic enhancement or suppression of the ultracold rotationally
resolved rate coefficients\cite{natcom2015,PRL2015,Hazra2015HO2}.
We note that a traditional GP calculation\cite{natcom2015,PRL2015,Hazra2015HO2}.
(which is computationally more feasible) on the ground adiabatic electronic state is not applicable for this system since the CI is located {\it below} the energy of the incident channel.

\subsection*{Rotationally Resolved Rate Coefficients as a function of collision energy}

\begin{figure}[h]
\includegraphics[scale=0.5]{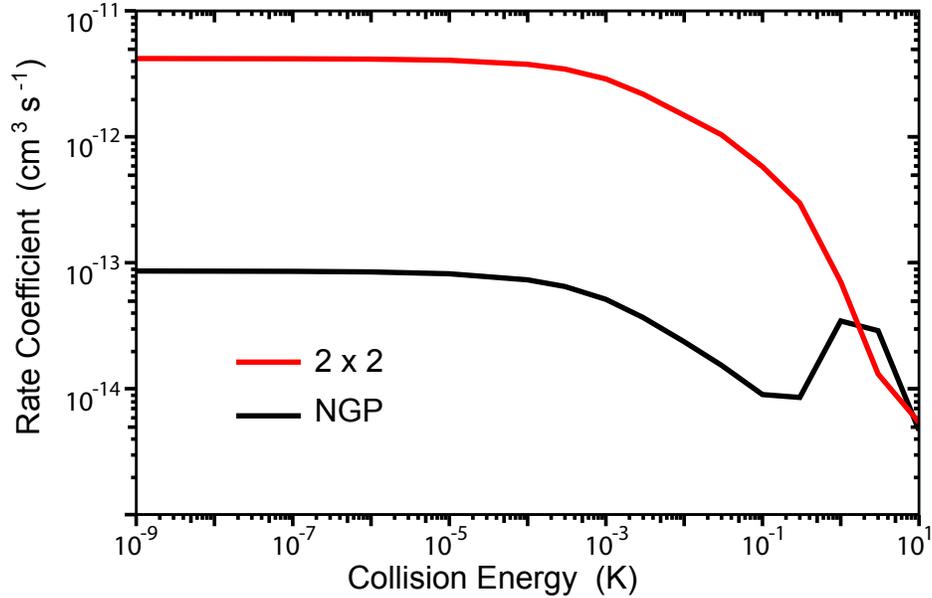}
\caption{
  \begin{small} { \textbf{Rotationally resolved rate coefficients.}
Rotationally resolved rate coefficients for the Li + LiNa($v=0$, $j=0$) $\to$ Li$_2$($v'=3$, $j'=5$) + Na reaction are plotted as a function of collision energy. The red and black curves are the rates computed using the coupled two-state diabatic ($2\times 2$) and single surface adiabatic (NGP) methods, respectively. The GP which is included in the diabatic $2\times 2$ calculations gives rise to constructive quantum interference and a significantly enhanced ultracold rate coefficient relative to the NGP calculation which ignores the GP.
}
\end{small}
}
\label{fig02}
\end{figure}

Figure \ref{fig02} plots a representative rotationally resolved rate coefficient for the Li + LiNa($v=0$, $j=0$) $\to$ Li$_2$($v'=3$, $j'=5$) + Na reaction as a function of collision energy from $1\,{\rm nK}$ to $10\,{\rm K}$.
Unless otherwise stated, all rate coefficients include the appropriate nuclear spin statistical factors of 2/3 and 1/3 for even and odd exchange symmetry (associated with the two identical $^6$Li nuclei), respectively.
At ultracold collision energies ($< 1\,{\rm mK}$), only a single partial wave (i.e., $l=0$ where $l$ is the orbital angular momentum of Li about LiNa) contributes to the collision and the rate coefficient becomes finite (often referred to as the Wigner regime)\cite{WignerLimit,balaThresh97}.
The specific values of the ultracold rate coefficients require exact quantum mechanical calculations on accurate PESs and are computationally demanding (see {\bf Materials and Methods} for details).
The red curve in Fig.~\ref{fig02} is from the coupled two-diabatic electronic states calculation ($2\times 2$) and the black curve is from the calculation on a single adiabatic ground electronic state which does {\it not} include the GP (denoted as NGP for No GP).
We see that in the ultracold limit the $2\times 2$ rate coefficient (red) is significantly enhanced ($\approx 50\times$) relative to the NGP one (black).
The enhancement is due to constructive quantum interference between the direct and looping contributions to the total scattering amplitude\cite{natcom2015}.
The GP associated with the conical intersection shown in Fig.~\ref{fig01} changes the sign of the interference term and hence the nature of the quantum interference from destructive to constructive and vice versa (for more details see the discussion and Eqs.~1 - 6 in
{\bf Supplementary Materials}).\cite{natcom2015,PRL2015,meadtruhlar79,mead80H3,berry84,kendrick2003,althorpe2005,zygelman2017,izmaylov2017,Xi2017,GPexp2018,GPexp2020}.
Furthermore, due to the unique properties of ultracold collisions, according to Levinson's theorem\cite{Levinson1953}
the scattering phase shifts preferentially approach an integral multiple of $\pi$.
Thus, the quantum interference often approaches its maximal values effectively turning the reaction on or off (i.e., a quantum switch!)\cite{natcom2015,PRL2015}.

The total rate coefficients summed over all final vibrational and rotational states of Li$_2$ are plotted in Fig.~S2.
The GP effects tend to wash out in the sum over final states so that the $2\times 2$ and NGP total ultracold rate coefficients are similar in magnitude (i.e., ${\rm K}_{\rm 2\times 2}/{\rm K}_{NGP}\approx 1.05$ at $E_c = 1.0\,{\rm nK}$).
Interestingly, both the $2\times 2$ and NGP ultracold rate coefficients lie below the universal value (i.e., ${\rm K}_{\rm 2\times 2}/{\rm K}_{\rm univ}\approx 0.89$ and
${\rm K}_{\rm NGP}/{\rm K}_{\rm univ}\approx 0.85$ at $E_c = 1.0\,{\rm nK}$).
The universal rate coefficient is computed using a simple one-dimensional model based only on the long-range $C_6$ potential along the reaction path and ignores all reflections\cite{C6_Hui2019}.
Thus, the smaller $2\times 2$ and NGP rates are most likely due to non-reactive (elastic) reflections that are included in the exact quantum mechanical calculations.
The sensitivity of the rate coefficients to the accuracy of the PES was also investigated.
Fig.~S3 plots the total $2\times 2$ and NGP rate coefficients as a function of a scaling parameter $\lambda$ for the PES.
The 3-body contribution to the PES is multiplied by $\lambda$ whereas the 2-body (pairwise) interaction potentials are left unchanged.
This ensures that the asymptotic energies and long-range interactions are unchanged and that only the effective depth of the Li$_2$Na PES is altered.
The range in $\lambda$ (i.e., $\pm 3\,\%$) was chosen to reflect the estimated uncertainty in the {\it ab initio} computed 3-body interaction PES.
Results for $\lambda=1$ correspond to the unscaled PES.
The NGP total rate coefficient oscillates between $2.16\times 10^{-10}$ and $3.74\times 10^{-10}\,{\rm cm}^3/{\rm s}$  (i.e., by $-20\,\%$ and $+38\,\%$ relative to the unscaled NGP rate coefficient).
The $2\times 2$ total rate coefficient oscillates between $2.46\times 10^{-10}$ and $4.49\times 10^{-10}\,{\rm cm}^3/{\rm s}$  (i.e., by $-13\,\%$ and $+58\,\%$ relative to the unscaled $2\times 2$ rate coefficient).
Interestingly, the effect of PES scaling on the rotationally resolved rate coefficients is much larger due to sudden changes in the nature of the quantum interference around the CI.
An example is plotted in Fig.~S4 for the Li$_2$($v'=3$, $j'=5$) + Na product state which shows large sudden enhancements or suppression in the rate coefficients as a function of $\lambda$ (see Figs. S5, S6, and Eqs. 1 - 6 for additional details).

\subsection*{Ultracold Rate Coefficient Distributions}

\begin{figure}[h]
\includegraphics[scale=0.35]{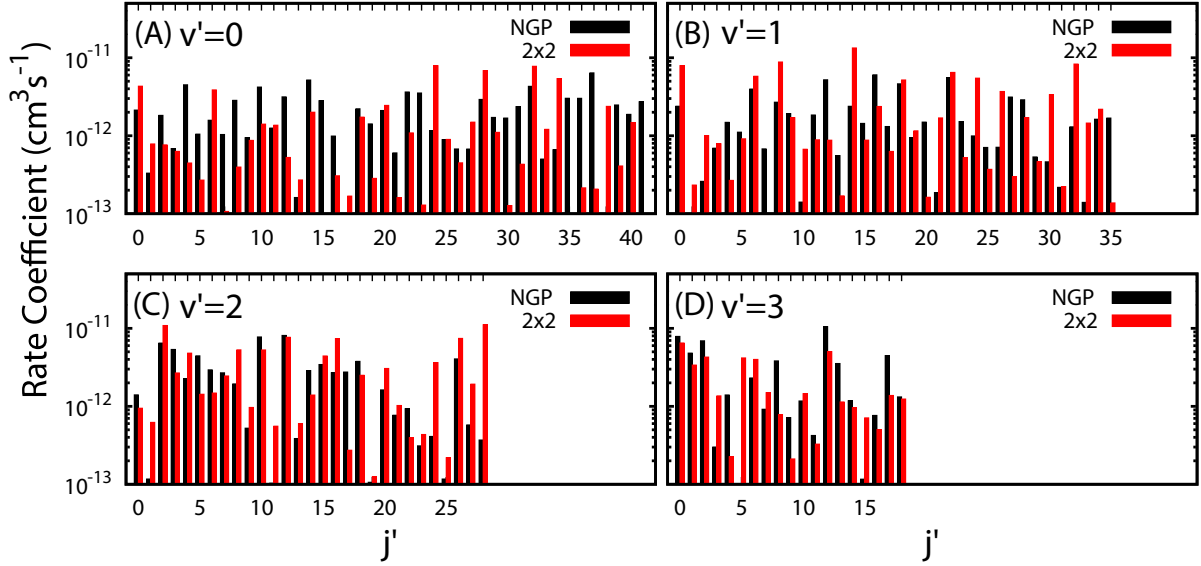}
\caption{
  \begin{small} { \textbf{Rotationally resolved rate coefficient distributions.}
        All of the rotationally resolved rate coefficients for the Li + LiNa($v=0$, $j=0$) $\to$ Li$_2$($v'$, $j'$) + Na reaction are plotted at the ultracold collision energy of $1.0\,{\rm nK}$. The red and black vertical bars are computed using the coupled two-state diabatic ($2\times 2$) and single surface adiabatic (NGP) methods, respectively. Panels (a), (b), (c) and (d) plot the distributions for $v'=0$, $1$, $2$ and $3$. Many of the diabatic ($2\times 2$) rate coefficients are significantly enhanced (suppressed) relative to the NGP rates due to constructive (destructive) quantum interference associated with the GP.
        }
\end{small}
}

\label{fig03}
\end{figure}

All of the rotationally resolved rate coefficients are plotted in Fig.~\ref{fig03} at the ultracold collision energy of $1\,{\rm nK}$ for each final
vibrational product state of Li$_2$ from $v'=0$ to $3$.
The red and black rate coefficients (vertical bars) correspond to the $2\times 2$ and NGP calculations, respectively.
Many of the $2\times 2$ rate coefficients are significantly enhanced or suppressed relative to the NGP rate coefficients.
As discussed above, this effect is due to the GP which is included in the $2\times 2$ calculations but not in the NGP calculations.
The sign change associated with the GP alters the nature of the quantum interference and hence the magnitude of the rate coefficients.
For $v'=0$ (panel A) particularly large GP effects are seen in the product rotational states $j'=4$, $7$, $15$, $23$, $30$, $35-37$ and $41$ for which the $2\times 2$ rate coefficients are suppressed relative to the NGP ones.
In contrast, the product rotational states for $j'=24$, $34$, and $38$ show significantly enhanced $2\times 2$ rates coefficients.
For $v'=1$ (panel B) notably suppressed $2\times 2$ rate coefficients are observed for the product rotational states $j'=12$, $20$, $27$, and $35$ whereas notably enhanced $2\times 2$ rate coefficients occur for $j'=14$, $21$, $24$, $26$, $30$, $32$, and $33$.
For $v'=2$ (panel C) notably suppressed $2\times 2$ rate coefficients are observed for the product rotational state $j'=17$ whereas notably enhanced $2\times 2$ rate coefficients occur for $j'=1$, $11$, $24$, $27$, and $28$.
Finally, for $v'=3$ (panel D) notably suppressed $2\times 2$ rate coefficients are observed for the product rotational states $j'=4$, $8$, $9$, $13$, and $17$ whereas notably enhanced $2\times 2$ rate coefficients occur for $j'=3$, $5$, and $15$.
In summary, the magnitude of the GP effect on the ultracold rotationally resolved rate coefficients varies significantly across all values of the product ro-vibrational states of Li$_2$($v'$, $j'$).

\begin{figure}[h]
\includegraphics[scale=0.5]{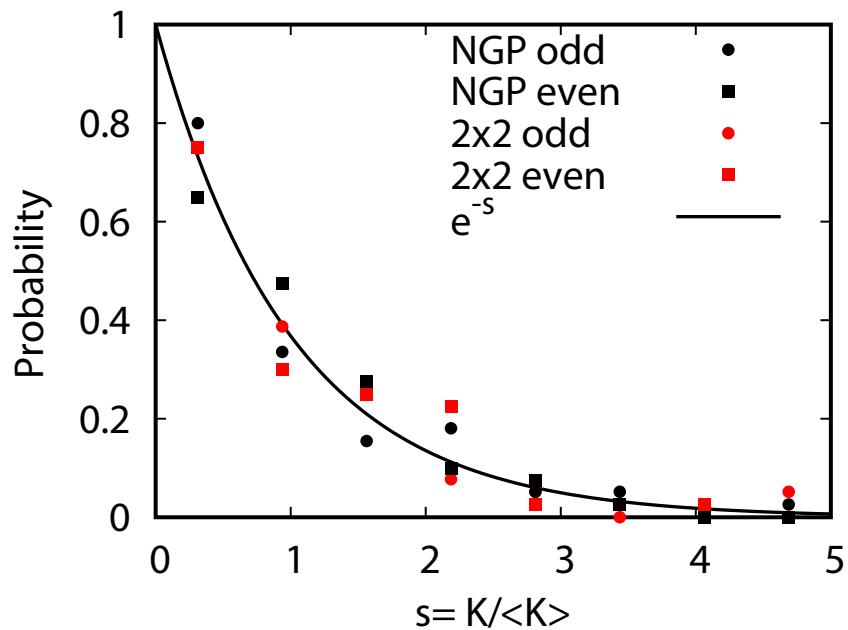}
\caption{
  \begin{small} { \textbf{Probability distributions.}
        The probability distributions for all of the rotationally resolved rate coefficients for the Li + LiNa($v=0$, $j=0$) $\to$ Li$_2$($v'$, $j'$) + Na reaction are plotted at the ultracold collision energy of $1.0\,{\rm nK}$. The distributions for the coupled two-state diabatic ($2\times 2$) and single surface adiabatic (NGP) methods are plotted in red and black, respectively. The results for even and odd exchange symmetry are plotted with circles and squares. The solid black curve is the Poisson distribution. All of the rate coefficient distributions are consistent with the Poisson distribution and are computed by binning the normalized rate coefficients $s={\rm K}/\langle {\rm K}\rangle$ where $\langle{\rm K}\rangle$ is the average rate coefficient.
        }
\end{small}
}
\label{fig04}
\end{figure}

Figure \ref{fig04} plots the normalized distributions $s=K/\langle K\rangle$ where $\langle K\rangle$ denotes the average value of the rate coefficients $K$ for a given data set.
The probability distributions are computed by binning the $K_{v'j'}$ into eight equally spaced intervals up to five times the average value.
Four normalized data sets are plotted. The red and black data points denote the $2\times 2$ and NGP rate coefficients, respectively.
The circles and squares correspond to the results of even and odd exchange symmetry.
The four data sets span all of the vibrational and rotational states shown in Fig.~\ref{fig03}.
For reference, the Poisson distribution ($e^{-s}$) is also plotted (solid black curve).
We see that on average all four data sets are consistent with the Poisson distribution.
Thus, a statistical analysis of the erratic looking rotational rate coefficient distributions of Fig.~\ref{fig03}
provides a unified description of all the results.
We note that the Poisson nature of the rotational distributions was also reported previously for the ultracold K + KRb reaction\cite{croftNatCom2017,croftPhysRev2017}.
This property appears to be very robust and is independent of the details of the PES and occurs for both the $2\times 2$ and NGP results.
For example, in Figs. S7 and S8 the Poisson distributions are plotted for 25 different values
of the PES scaling parameter $\lambda$ for each exchange symmetry even and odd, respectively.
The collective set of $100$ distributions are consistent with the Poisson distribution.
The K + KRb results\cite{croftNatCom2017,croftPhysRev2017} together with the present work confirms what appears to be a universal property of ultracold chemical reactions
with a potential well supporting long-lived complex formation: the rotationally resolved rate coefficient probability distributions are Poissonian.

{\color{red}\section*{Discussion}}
Many ultracold chemical reactions under active experimental investigation, such as Li + LiNa $\to$ Li$_2$ + Na, K + NaK $\to$ K$_2$ + Na, KRb + KRb $\to$ K$_2$ + Rb$_2$,
and NaRb + NaRb $\to$ Na$_2$ + Rb$_2$\cite{Rvachov2017,Rui_ExpNaK2017,Ye_ExpNaRb2018,Hu_KRb2019}
have a barrierless reaction pathway and a deep attractive potential well.
In addition, they also exhibit a CI between the ground and first excited electronic states in the interaction region.
This CI is energetically accessible even for ultracold collisions involving reactant diatomic molecules
in their ground ro-vibrational state (e.g., LiNa($v=0$, $j=0$)).
Thus, an exact quantum mechanical calculation is required which includes both electronic states using accurate {\it ab initio} PESs\cite{Kendrick2018nonad}.
To the authors' knowledge, the first non-adiabatic calculations of this kind are reported in this work for the ultracold Li + LiNa($v=0$, $j=0$) $\to$ Li$_2$($v'$, $j'$) + Na
reaction.

Two reaction pathways (direct and looping) which encircle the CI contribute to the ultracold rate coefficients for the Li + LiNa reaction
and the resulting quantum interference between these two pathways can be constructive or destructive.
Due to the unique properties of ultracold collisions, the quantum interference often approaches its maximal values
which leads to a significantly enhanced or suppressed rate coefficient (i.e., the reaction is effectively turned on or off).
Furthermore, the GP associated with the CI changes the sign on the interference term which
reverses the nature of the quantum interference.
Thus, a non-adiabatic calculation which includes the excited electronic state and its associated GP is crucial for obtaining the correct theoretical prediction of the rate coefficients.
A conventional Born-Oppenheimer calculation based on a single adiabatic ground electronic state PES will give the opposite (incorrect) prediction whenever significant quantum interference occurs.
The novel quantum interference mechanism associated with ultracold collisions represents a realization of a molecular quantum switch.
The large dynamic range of this quantum switch might be exploited by experimentalists to control the reaction outcome via 
the application of external fields and/or the selection of a particular initial quantum state\cite{natcom2015,PRL2015,Hazra2015HO2}.

The large quantum interference effects observed in the rotationally resolved rate coefficients mostly cancel out in the total rate coefficient summed over all product states.
The total ultracold rate coefficients for the non-adiabatic and adiabatic calculations differ by only $5\,\%$.
Interestingly, the ultracold rate coefficients from both sets of calculations lie about $10$ to $15\,\%$ below the universal value based on a
simple one-dimensional long-range (C$_6$) potential. This non-universal behavior suggests that non-reactive (i.e., elastic) reflections are significant.
In contrast, excellent agreement between exact quantum dynamics calculations and a universal model was reported for the K + KRb reaction\cite{croftNatCom2017}.

The rotationally resolved rate coefficient distributions are also shown to exhibit Poisson behavior.
The ${\bf S}$ matrix for open chaotic quantum systems obeys the statistics of unitary symmetric random matrices,
one of which is the Poisson law behavior of the squares of off-diagonal matrix elements\cite{Blumel88,Dyson62}.
Since state-to-state rates are directly proportional to the square of the corresponding ${\bf S}$ matrix element, this Poisson law
behavior follows directly from the underlying classically chaotic motion of the reaction\cite{Honvault2000}.
Chaotic classical trajectories are extremely complicated and tangled for reactions with long-lived intermediate complexes,
as such these results show that the ultracold LiNa + Li reaction proceeds via complex formation.
Such intermediate complexes can be observed experimentally using a combination of mass spectrometry and velocity map imaging,
as was recently demonstrated for the ultracold KRb + KRb $\to$ K$_2$ + Rb$_2$ reaction\cite{Hu_KRb2019}.
As shown explicitly in this work for the first time, the Poisson nature of these rotational distributions is robust to variations in the PES,
occurs for different chemical systems (i.e., both light Li$_2$Na and heavy KKRb\cite{croftNatCom2017})
and theoretical methods (i.e., both non-adiabatic ($2\times 2)$ and adiabatic (NGP)).
The robust and universal nature of the Poisson ultracold rotational rate coefficient distributions makes this property an ideal experimental observable.

We hope that the theoretical results presented in this work will help stimulate new experimental and theoretical studies into the intriguing
ultracold energy regime. The unique properties of ultracold collisions are still largely unexplored.
Ultracold molecules continue to show exceptional promise for future technological applications in quantum control, sensing and precision measurements.

{\color{red} \section*{Materials and Methods}}
\noindent
{\bf {Potential Energy Surfaces of LiNaLi}.}
Accurate and complete information on PESs of the LiNaLi collisional complex are absent in the literature and their computation required substantial effort due to the complexity of the multi-electron open-shell systems.
Our electronic structure calculations have been carried out with the MOLPRO program package \cite{molpro}. Core electron shells of Li and Na  are described by the Stuttgart/Cologne energy-consistent, single-valence electron, relativistic pseudo-potentials (ECPs), ECP2SDF and 
ECP10SDF\cite{fuentealba1983psd},
leaving only three valence electrons in the active space for explicit treatment.
The polarization of the effective cores and  residual core-valence correlations are modeled via the {\it l}-independent core polarization potential 
(CPP) with M{\"u}ller-Meyer damping functions \cite{muller1984treatment}. The CPP parameters, {i.e.} the static dipole polarizabilities of the atomic cores, $\alpha^{+}_{\rm c}$, are taken from Ref.~\cite{mitroy2010theory} and the cutoff functions with exponents $0.95$ a.u. and $0.82$ a.u. for Li and Na, respectively, are employed. Here, a.u. stands for atomic unit. Basis sets from Ref.~\cite{zuchowski2010reactions}  describe the three valence electrons, specifically,  uncontracted  $sp$ basis sets  augmented by additional $s$, $p$, $d$ and $f$ polarization functions are used for both Li and Na. The multi-configurational 
self-consistent field (MCSCF) \cite{werner1985seconda,werner1985_2} method is first used to obtain configuration state functions (CSFs).
An MRCI calculation is then performed using a large active space constructed from the CSFs, giving the  three-dimensional adiabatic surfaces of the two lowest energy states for LiNaLi, $V_1$ and $V_2$ as functions of the three bond-lengths.  Nonadiabatic coupling matrix elements between these two electronic surfaces are computed at the same level of MRCI  theory with the numerical finite differential method (DDR procedure). For use in the reactive scattering calculations the
non-adiabatic coupling function is spatially integrated to generate the three-dimensional mixing angle $\beta$ \cite{Domcke2004}.
Finally,  fitted global full-dimensional PESs were constructed from the {\it ab~initio} energies using the reproducing kernel Hilbert space (RKHS) technique \cite{Ho1996, Unke2017}.

\vspace {5mm}
\noindent
{\bf Non-adiabatic Quantum Dynamics}. 
The non-adiabatic quantum dynamics calculations solve the time-independent two-state ($2\times 2$) diabatic Schr\"odinger equation for the nuclear motion given by\cite{Kendrick2018nonad}
\begin{equation}
\left [ 
\left (
\begin{array}{cc}
{\hat T} & 0 \\
0 & {\hat T} \\
\end{array}
\right )
+  \left (
\begin{array}{cc}
{\tilde V}_{11} & {\tilde V}_{12}\\
{\tilde V}_{21} & {\tilde V}_{22}\\
\end{array}
\right )
\right ]\,\left (
\begin{array}{c}
{\tilde \psi}_1\\
{\tilde \psi}_2\\
\end{array}
\right ) =  E\,
\left (
\begin{array}{c}
{\tilde \psi}_1\\
{\tilde \psi}_2\\
\end{array}
\right )
\label{Diab2x2eq}
\end{equation}
where the first term in brackets in Eq. \ref{Diab2x2eq} is the diabatic kinetic energy operator 
for the nuclear motion with matrix elements ${\hat T}={-\hbar^2\over 2\mu}\,\nabla^2$ where
$\nabla$ denotes the derivatives with respect to the six nuclear coordinates (three bond lengths and three Euler angles)
relative to the center of mass and $\mu$ is the three-body reduced mass.
The second term is the diabatic potential matrix $\tilde{\bf V}$ which is a function of the three bond lengths with matrix elements given by
\begin{align}
{\tilde V}_{11}&= V_1\,\cos^2\beta  + V_2\,\sin^2\beta\, , \label{V11eq}\\
{\tilde V}_{22}&= V_2\,\cos^2\beta + V_1\,\sin^2\beta\, ,  \label{V22eq}\\
{\tilde V}_{12}&= {\tilde V}_{21} = (V_2 - V_1)\,\cos\beta\,\sin\beta , \label{V12eq}
\end{align}
where $V_1$ and $V_2$ are the adiabatic PESs and $\beta$ is their mixing angle as discussed above.
In contrast to the $2\times 2$ diabatic Schr\"odinger Eq. \ref{Diab2x2eq}, the conventional Born-Oppenheimer (NGP) quantum dynamics calculations solve the adiabatic single surface Schr\"odinger equation
\begin{equation}
\Bigr [- {\hbar^2\over 2\,\mu}\,\nabla^2 + V_1({\rm x})\Bigl ] \,\psi_1({\bf x}) = E\,\psi_1({\bf x}) \, .
\label{BOeq}  
\end{equation}

The quantum dynamics calculations use Adiabatically adjusting Principal axis Hyperspherical (APH) coordinates
in the interaction region and Delves hyperspherical coordinates in the long-range asymptotic region\cite{packparker87,Kendrick99,Kendrick2018nonad}.
The hyperradius $\rho$ is common to both coordinate systems which facilitates the coordinate transformation from the APH to Delves at an intermediate value of $\rho_m$ (determined by numerical convergence studies).
In the interaction region, the two-dimensional (2D) surface function Hamiltonian matrix is diagonalized on a discrete grid in $\rho$ ($144$ logarithmically spaced points were used between $\rho_i=6.0\,{\rm a}_{\rm 0}$
and $\rho_m=33.0\,{\rm a}_{\rm 0}$). The 2D basis functions consist of a hybrid FBR (Finite Basis Representation) in $\phi$ and DVR (Discrete Variable Representation) in $\theta$\cite{light85dvr_sdt,Kendrick99,Kendrick2018nonad}.
The size of the FBR and DVR varies with $\rho$ and is determined from numerical convergence studies. The size of the 2D Hamiltonian matrix is dramatically reduced by using SDT (Sequential Diagonalization Truncation)\cite{light85dvr_sdt}.
An efficient numerical eigensolver (PARPACK) is used to numerically diagonalize the sparse 2D Hamiltonian matrix\cite{arpack}.
For the zero total angular momentum ($J=0$) studied in this work, the matrix dimension varied between approximately $10\,000$ for small $\rho$ to $2\,500$ for large $\rho$.
The set of 2D eigensolutions form a basis for the one-dimensional coupled-channel propagation in $\rho$.
A log-derivative propagation technique is used to propagate a matrix of solutions (the log-derivative matrix) from $\rho=\rho_i$ to $\rho=\rho_m$.
The number of coupled channels propagated in this work was $820$ in the APH region and $500$ in the Delves region for each exchange symmetry even or odd.
The Delves functions for each diatomic arrangement channel consist of ro-vibrational wave functions computed numerically using a one-dimensional Numerov propagator for the vibrational motion
and a set of analytic spherical harmonics for the rotational part.
The log-derivative matrix is transformed from the APH to Delves coordinates at $\rho=\rho_m$ using the overlap matrix between the APH and Delves wave functions.
The log-derivative propagation is then continued using the Delves ro-vibrational basis across $482$ uniformly spaced $\rho$ values to the final asymptotic $\rho_f= 144.6\,{\rm a}_{\rm 0}$.
At the final value of $\rho=\rho_f$, the overlap matrix between the Delves functions and Jacobi basis functions is computed which enables the evaluation of the scattering ${\bf S}$ matrix\cite{packparker87}.
Once the ${\bf S}$ matrix is computed, cross sections $\sigma_{fi}$ and rate coefficients $K_{fi}=v\,\sigma_{fi}$ (where $v$ is the relative collision velocity) can be computed using standard expressions\cite{packparker87}.
We note that the $f,i$ denote the collective final and initial quantum numbers of the diatomic products and reactants (i.e., $f=(\tau',v',j')$ and $i=(\tau,v,j)$
where $\tau'$ and $\tau$ denote the diatomic arrangement channel Li$_2$ or LiNa), respectively.

\pagebreak

{\color{red} \section*{Supplementary Materials}}
Supplementary material for this article is available at http://adavances.sciencemag.org/xxx

\clearpage
{\color{red}\section*{References and Notes}}

\clearpage
\subsection*{Acknowledgements}
\textbf{Funding}: B.K.K. acknowledges that part of this work was done under the auspices of the US Department of Energy under Project No. 20170221ER
of the Laboratory Directed Research and Development Program at Los Alamos National Laboratory.
Los Alamos National Laboratory is operated by Triad National Security, LLC, for the National Nuclear Security
Administration of the U.S. Department of Energy (contract No. 89233218CNA000001).
S.K. acknowledges support from the Army Research Office
Grant No. W911NF-17-1-0563, the NSF Grant Nos. PHY-1619788 and PHY-1908634. 
J.F.E.C. acknowledges that part of this work was supported by the Marsden Fund of New Zealand (Contract No. UOO1923) and gratefully acknowledges support from the Dodd-Walls Centre for Photonic and Quantum Technologies.
N.B. acknowledges partial support from NSF grant No. PHY-1806334.
\textbf{Author contributions:} B.K.K. performed the majority of the numerical quantum dynamics
calculations, scattering analysis and wrote the manuscript.
S.K., M.L. and H.L. performed the electronic structure calculations of the PES and universal model calculations.
J.F.E.C. performed the statistical analysis and initial numerical quantum dynamics calculations with assistance from N.B. and B.K.K.
N.B. and S.K. conceived the research project and all authors reviewed and commented on the manuscript.
\textbf{Competing Interests:} The authors declare that they 
have no competing interests.
\textbf{Data and materials availability:} All data needed to 
evaluate the conclusions in the paper are present in the paper and/or the Supplementary 
Materials. Additional data related to this paper may be requested from the authors.

\clearpage
\section*{Figures}

\makeatletter
\renewcommand{\fnum@figure}{\textbf{Fig.~\thefigure}}
\makeatother

\clearpage


%





\setcounter{figure}{0}
\renewcommand\thefigure{{\arabic{figure}}}
\section*{Supplementary Materials}

\begin{figure}[h]
\includegraphics[scale=0.65]{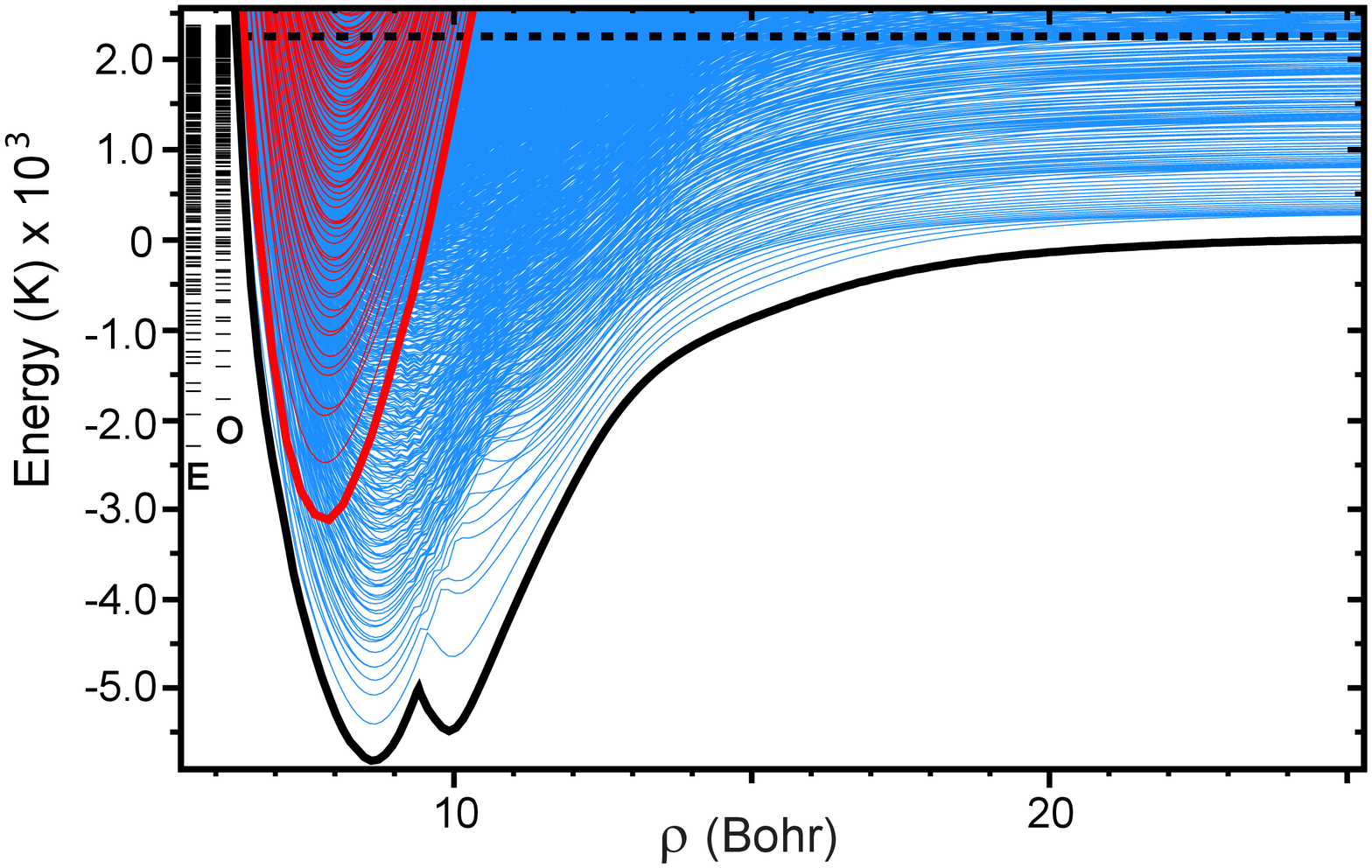}
\caption{
  \begin{small} {\textbf{Surface function energies.}
The APH surface function energies are plotted as a function of the hyperradius $\rho$. The blue and red adiabats are computed on the electronic adiabatic ground ($V_1$) and excited ($V_2$) states, respectively.  The thick black and red curves plot the minimum energy of the ground and excited adiabatic electronic states at each $\rho$, respectively. The horizontal black dashed line denotes the total energy of the Li + LiNa($v=0$, $j=0$) reactant ($2\,228{\rm K}$). The blue adiabats on the right edge of the plot (i.e., large $\rho$) which lie below the black horizontal line correlate to the asymptotic ro-vibrational energies of the product Li$_2$($v'$, $j'$) + Na. The series of energy levels labeled by the "E" (even exchange symmetry) and "O" (odd exchange symmetry) are the three-dimensional vibrational energies computed on the excited electronic state (i.e., "cone states"). All energies are relative to the minimum energy of the asymptotic adiabatic ground electronic state of Li$_2$ + Na.
    }
\end{small}
}
\label{figS1}
\end{figure}

Figure S\ref{figS1} plots the adiabatic surface function energies for the ground (blue) and excited (red) electronic states. These surface function energies were computed for each electronic state separately (i.e., an uncoupled single adiabatic electronic state calculation). The black and red thick solid curves plot, respectively, the minimum energies of the ground and excited adiabatic electronic state PESs as a function of $\rho$.
Asymptotically for large $\rho$, the ground state adiabatic surface function energies (blue) approach the diatomic Li$_2$($v'$, $j'$) and LiNa($v$, $j$) rovibrational energies.
The dashed horizontal black line denotes the energy of the Li + LiNa($v=0$, $j=0$) reactant channel ($2\,228\,{\rm K}$).
The excited electronic state is energetically open in the interaction region (small $\rho$) where the red curves drop below the black dashed horizontal line.
The vertical series of short horizontal lines on the left edge of the plot denote the bound state energies of the cone states which are localized inside the cone ($V_2$) of the excited adiabatic electronic state (see Fig.~1 of the paper).
The E and O label the bound states of even and odd exchange symmetry, respectively.

\begin{figure}[h]
\includegraphics[scale=0.5]{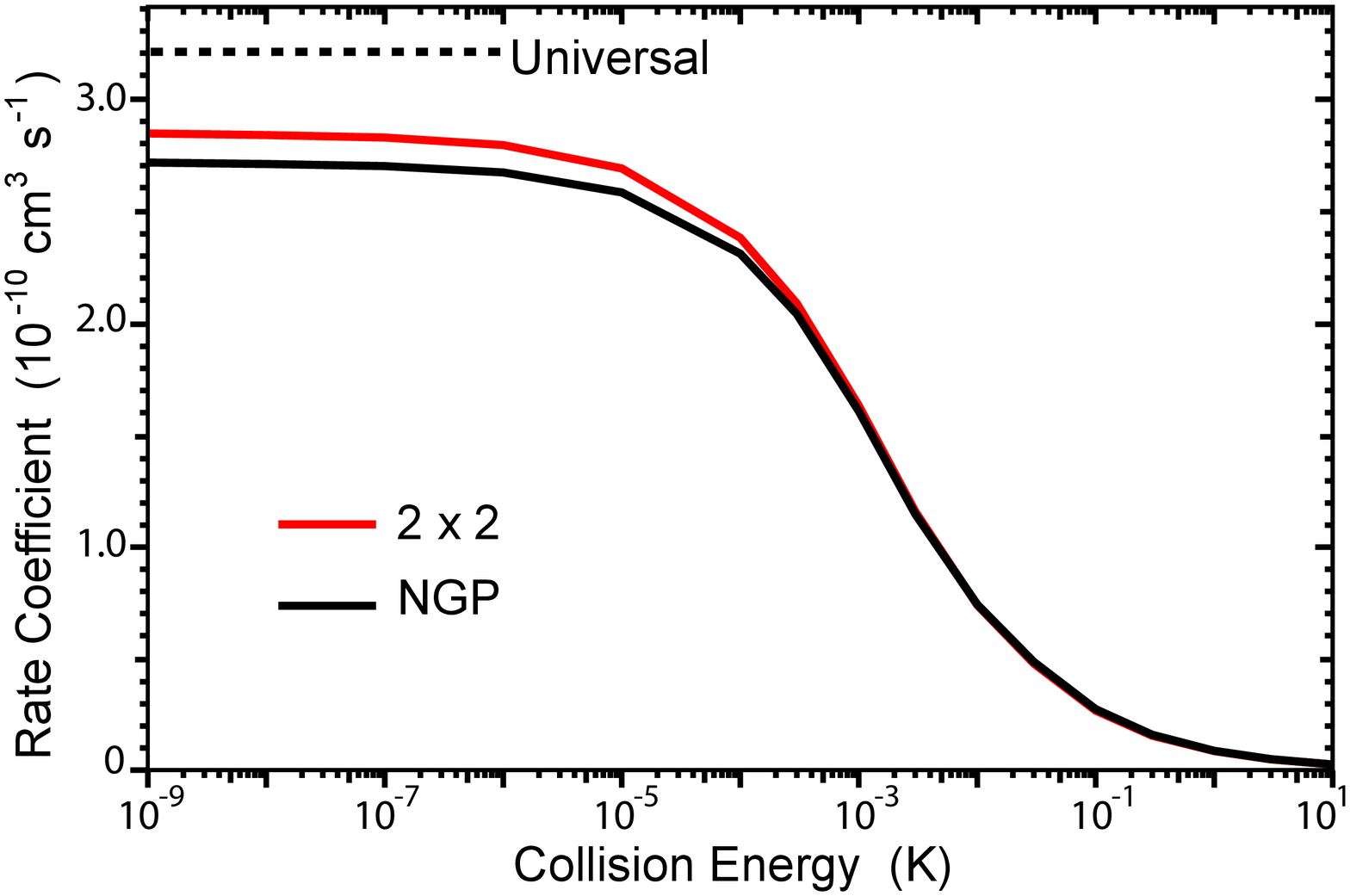}
\caption{
  \begin{small} {\textbf{Total rate coefficient.}
The total rate coefficient is plotted as function of collision energy for the Li + LiNa($v=0$, $j=0$) $\to$ Li$_2$ + Na reaction. The red and black curves correspond to the coupled two-state diabatic ($2\times 2$) and single surface adiabatic (NGP) calculations, respectively. The horizontal black dashed line denotes the ultracold rate coefficient computed using a universal model.         
        }
\end{small}
}
\label{figS2}
\end{figure}

Figure S\ref{figS2} plots the total rate coefficient for the Li + LiNa($v=0$, $j=0$) $\to$ Li$_2$ + Na reaction as a function of collision energy.
The total rates are computed by summing over the rate coefficients for all final product states Li$_2$($v'$, $j'$) with the even and odd exchange symmetry contributions (i.e., even and odd $j'$)
weighted by the appropriate nuclear spin statistical factors.
The red and black curves correspond to the $2\times 2$ and NGP calculations, respectively.
The horizontal black dashed line is the ultracold rate coefficient computed using a simple one-dimensional universal model based on just a long-range C$_6$ potential.

\begin{figure}[h]
\includegraphics[scale=0.5]{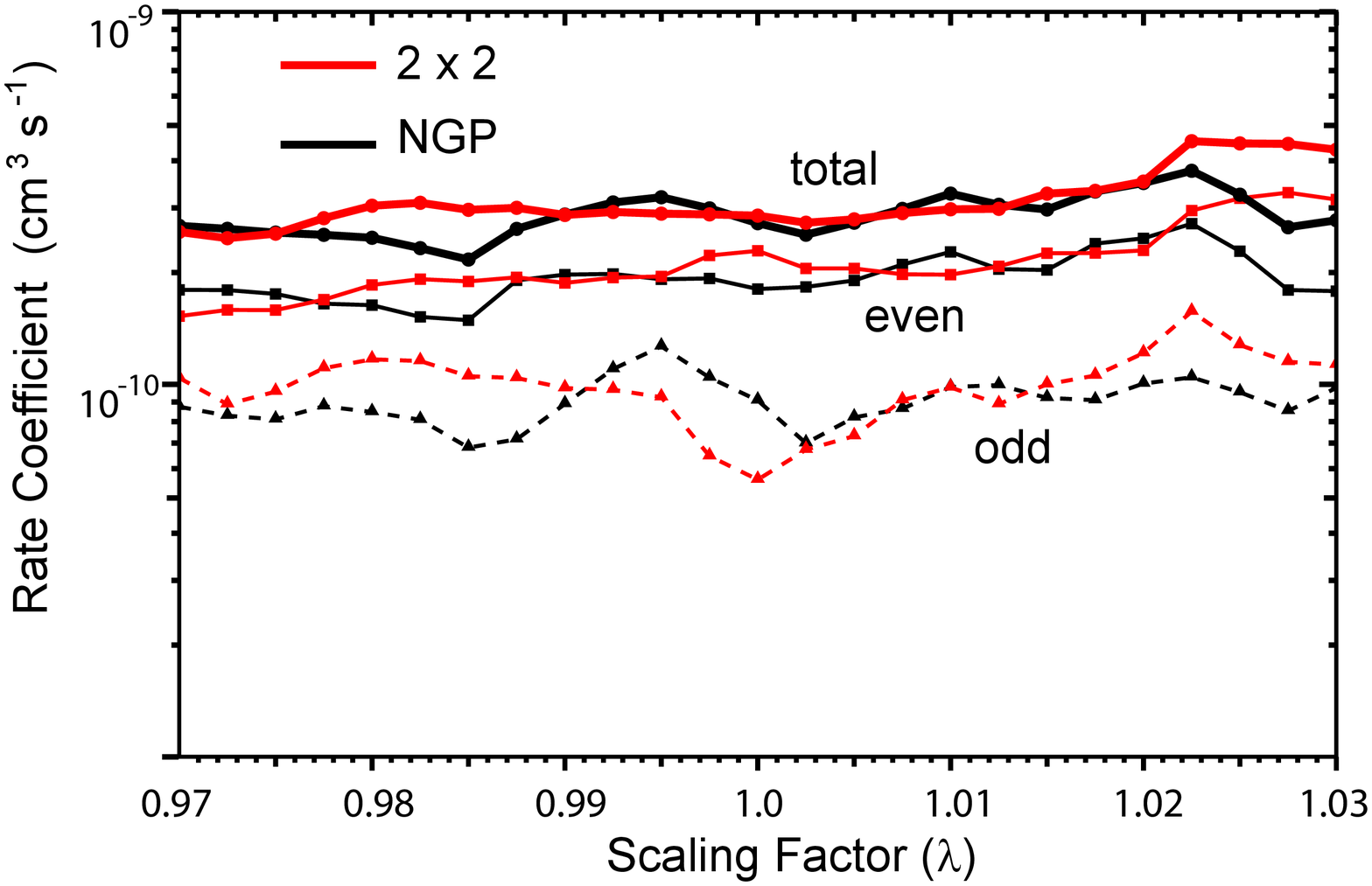}
\caption{
  \begin{small} {\textbf{Three-body potential scaling effects on the total rate coefficient.}
The total rate coefficients for the Li + LiNa($v=0$, $j=0$) $\to$ Li$_2$ + Na reaction are plotted as a function of the 3-body potential scaling parameter at the ultracold collision energy of $1\,{\rm nK}$.
The red and black data correspond to the coupled two-state diabatic ($2\times 2$) and single surface adiabatic (NGP) calculations, respectively.
The large circular data points are the total rate coefficients computed by adding the statistically weighted rate coefficients for even exchange symmetry (squares) and odd exchange symmetry (triangles).
To guide the eye, the data points for the total, even, and odd rate coefficients are connected by thick lines, thin lines and dashed lines, respectively.        
        }
\end{small}
}
\label{figS3}
\end{figure}

Figure S\ref{figS3} plots the ultracold total rate coefficient for the Li + LiNa($v=0$, $j=0$) $\to$ Li$_2$ + Na reaction as a function of the three-body potential scaling factor ($\lambda$).
The {\it ab initio} computed three-body contribution to the potential energy surface (PES) was scaled by $\lambda$ but the pairwise two-body potentials were left unchanged.
The scaling studies span a range of $25$ equally spaced scaling factors between $\pm 3\,\%$ (i.e. $\lambda=0.97$ to $\lambda=1.03$).
A total of $100=4\times 25$ calculations (which include $\lambda=1$ no scaling) were performed for the $2\times 2$ and NGP methods and each exchange symmetry even and odd.
The $2\times 2$ and NGP results are presented in red and black, respectively.
The total, even, and odd rate coefficients are plotted with solid circles, squares, and triangles, respectively.

\begin{figure}[h]
\includegraphics[scale=0.5]{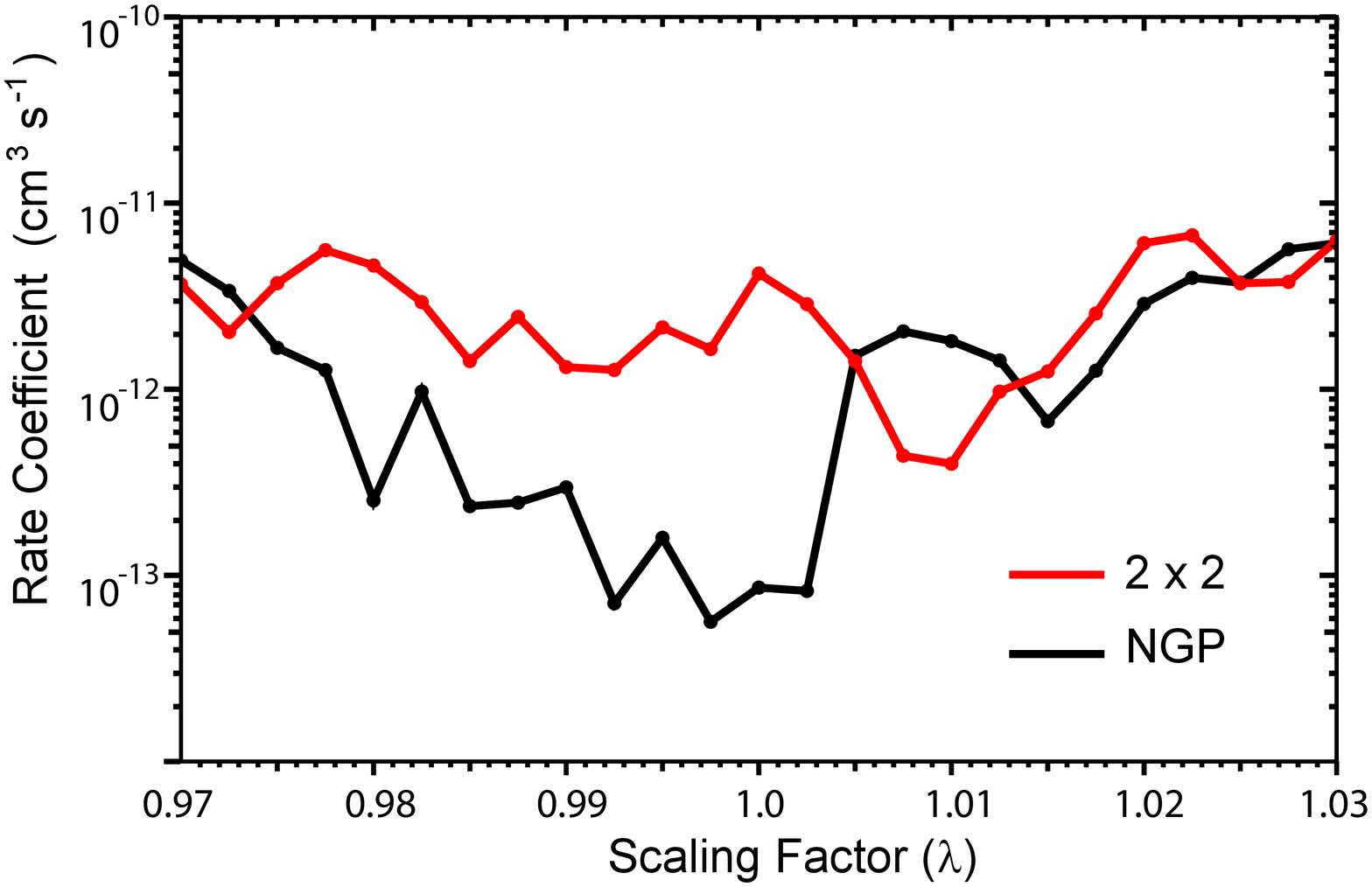}
\caption{
  \begin{small} {\textbf{Three-body potential scaling effects on a rotationally resolved rate coefficient.}
Rotationally resolved rate coefficients for the Li + LiNa($v=0$, $j=0$) $\to$ Li$_2$($v'=3$, $j'=5$) + Na reaction are plotted as a function of the 3-body potential scaling parameter at the ultracold collision energy of $1\,{\rm nK}$. The red and black data correspond to the coupled two-state diabatic ($2\times 2$) and single surface adiabatic (NGP) calculations, respectively.
      }
\end{small}
}
\label{figS4}
\end{figure}

Figure S\ref{figS4} plots the ultracold rotationally resolved rate coefficient for the Li + LiNa($v=0$, $j=0$) $\to$ Li$_2$($v'=3$, $j'=5$) + Na reaction as a function of the three-body potential scaling factor ($\lambda$).
The $2\times 2$ and NGP results are presented in red and black, respectively.
As the scaling factor increases or decreases away from $\lambda=1$ (i.e., no scaling), dramatic and often sudden changes in both the $2\times 2$ and NGP rate coefficients are observed.
These large changes are due to the change in nature of the quantum interference from constructive to destructive and vice versa.
The quantum interference occurs between the direct and looping contributions to the total scattering amplitude.
The direct pathway proceeds directly from reactants to products through the Li$_2$Na potential well whereas the looping pathway ``loops around'' and encircles the conical intersection (see Fig.~1 in the paper).

Let ${\tilde f}_{\rm direct}=f_{\rm direct}\,e^{i\,\delta_{\rm direct}}$ and ${\tilde f}_{\rm loop}=f_{\rm loop}\,e^{i\,\delta_{\rm loop}}$ denote the complex scattering amplitudes associated with the direct and looping pathways, respectively.
The $f_{\rm direct}$ and $f_{\rm loop}$ are the real valued magnitudes and the $\delta_{\rm direct}$ and $\delta_{\rm loop}$ are the real valued phases.
The total scattering amplitude is the sum of the two
\begin{equation}
{\tilde f}_{\rm total} = {1\over\sqrt{2}}( {\tilde f}_{\rm direct} + {\tilde f}_{\rm loop})\, .
\label{ftotal}
\end{equation}
The cross sections and rate coefficients are computed from the modulus of the total scattering amplitude given by
\begin{equation}
\vert {\tilde f}_{\rm total}\vert^2 = {1\over 2}( f^2_{\rm direct} + f^2_{\rm loop} + 2\,f_{\rm direct}\,f_{\rm loop}\,\cos\Delta)\, ,
\label{ftotal2}
\end{equation}  
where the relative phase $\Delta= \delta_{\rm loop} - \delta_{\rm direct}$.
The third term on the right hand side of Eq.~\ref{ftotal2} is the interference term.
If the magnitudes of the looping and direct scattering amplitudes are comparable in magnitude: $f_{\rm direct}\approx f_{\rm loop} = f$, then
Eq.~\ref{ftotal2} becomes
\begin{equation}
\vert {\tilde f}_{\rm total}\vert^2 \approx f^2\,( 1 + \cos\Delta)\, .
\label{ftotal2_red}
\end{equation}  
At ultracold collision energies the scattering phase shifts $\delta_{\rm loop}$ and $\delta_{\rm direct}$ have a propensity to
approach an integral multiple of $\pi$ (i.e., Levinson's theorem): $\delta_{\rm direct} \approx \pi\,n_{\rm direct}$ and $\delta_{\rm loop} \approx \pi\,n_{\rm loop}$
where the $n_{\rm direct}$ and $n_{\rm loop}$ are integers.
Thus the relative phase also approaches an integral multiple of $\pi$: $\Delta\approx \pi\,(n_{\rm loop} - n_{\rm direct}) \approx n\,\pi$ where the integer $n=n_{\rm loop} - n_{\rm direct}$.
This implies that the interference term $\cos\Delta$ in Eq.~\ref{ftotal2_red} approaches: $\cos\Delta \approx 1$ or $\cos\Delta \approx 0$ for even and odd values of $n$, respectively.
From Levinson's theorem we know that the integers $n_{\rm direct}$ and $n_{\rm loop}$ denote the number of bound states supported along the direct and looping reaction pathways, respectively.
Thus, the integer $n$ is the relative number of bound states between these two pathways which in general is either even or odd.
In practice, we never need to explicitly compute these integers (or $\Delta$) but can infer the even or odd nature of $n$ from the scattering amplitudes.
Equations \ref{ftotal} - \ref{ftotal2_red} are valid for both the $2\times 2$ and NGP calculations.
However, the $2\times 2$ calculations include the GP which gives rise to an additional sign change or $\pi$ phase shift in the scattering amplitude for the looping pathway (relative to the NGP one).
That is, $\delta^{2\times 2}_{\rm loop} =  \delta^{NGP}_{\rm loop} + \pi + \epsilon$  where $\epsilon$ represents other differences
(due to non-adiabatic couplings) between the NGP and $2\times 2$ scattering phase shifts.
The GP phase shift is defined as $\delta^{GP}_{\rm loop} = \delta^{NGP}_{\rm loop} + \pi$, so that we can also write $\delta^{2\times 2}_{\rm loop} =  \delta^{GP}_{\rm loop} + \epsilon$.
If other non-adiabatic effects are small, then $\epsilon \approx 0$ and we obtain $\delta^{2\times 2}_{\rm loop} \approx  \delta^{GP}_{\rm loop}$ (and similarly $\delta^{NGP}_{\rm direct}=\delta^{GP}_{\rm direct}\approx\delta^{2\times 2}_{\rm direct}$).

Substituting the above expressions for the looping and direct phase shifts for $\delta^{2\times 2}$ and $\delta^{NGP}$ into the $\Delta$ in Eq.~\ref{ftotal2_red} (and ignoring $\epsilon$), we obtain
\begin{equation}
\vert {\tilde f}^{NGP/2\times 2}_{\rm total}\vert^2 \approx f^2\,( 1 \pm \cos(n\,\pi))\, ,
\label{ftotal2_cosn}
\end{equation}  
where the $+$ sign corresponds to NGP and the $-$ sign corresponds to $2\times 2$ (or GP).
If $n$ is an even integer, then Eq.~\ref{ftotal2_cosn} gives
\begin{align}
  \vert {\tilde f}^{NGP}_{\rm total}\vert^2 &\approx 2\,f^2 &(even\ n)\cr
  \vert {\tilde f}^{2\times 2}_{\rm total}\vert^2 &\approx 0  &(even\ n)\, .
\label{ftotal2_NGPeven}
\end{align}
If $n$ is an odd integer, then Eq.~\ref{ftotal2_cosn} gives
\begin{align}
  \vert {\tilde f}^{NGP}_{\rm total}\vert^2 &\approx 0\, &(odd\ n)\cr
  \vert {\tilde f}^{2\times 2}_{\rm total}\vert^2 &\approx 2\,f^2 &(odd\ n)\, .
\label{ftotal2_NGPodd}
\end{align}
In the ultracold limit, equations \ref{ftotal2_NGPeven} and \ref{ftotal2_NGPodd} show that
the quantum interference can approach its maximal values of $2\,f^2$ (constructive) or zero (destructive).
The constructive or destructive nature of the interference is always opposite for the NGP and $2\times 2$ calculations
and can reverse if the integer $n$ changes from being even to odd or vice versa.
We note that the maximal values given by Eqs.~\ref{ftotal2_NGPeven} and \ref{ftotal2_NGPodd} are not
fully realized in practice since the direct and looping scattering amplitudes are not exactly equal in magnitude and their
scattering phase shifts are not exactly an integer multiple of $\pi$.
However, as discussed below, these limiting cases are useful for understanding and interpreting the large differences observed between the $2\times 2$ and NGP results.

\begin{figure}[h]
\includegraphics[scale=0.5]{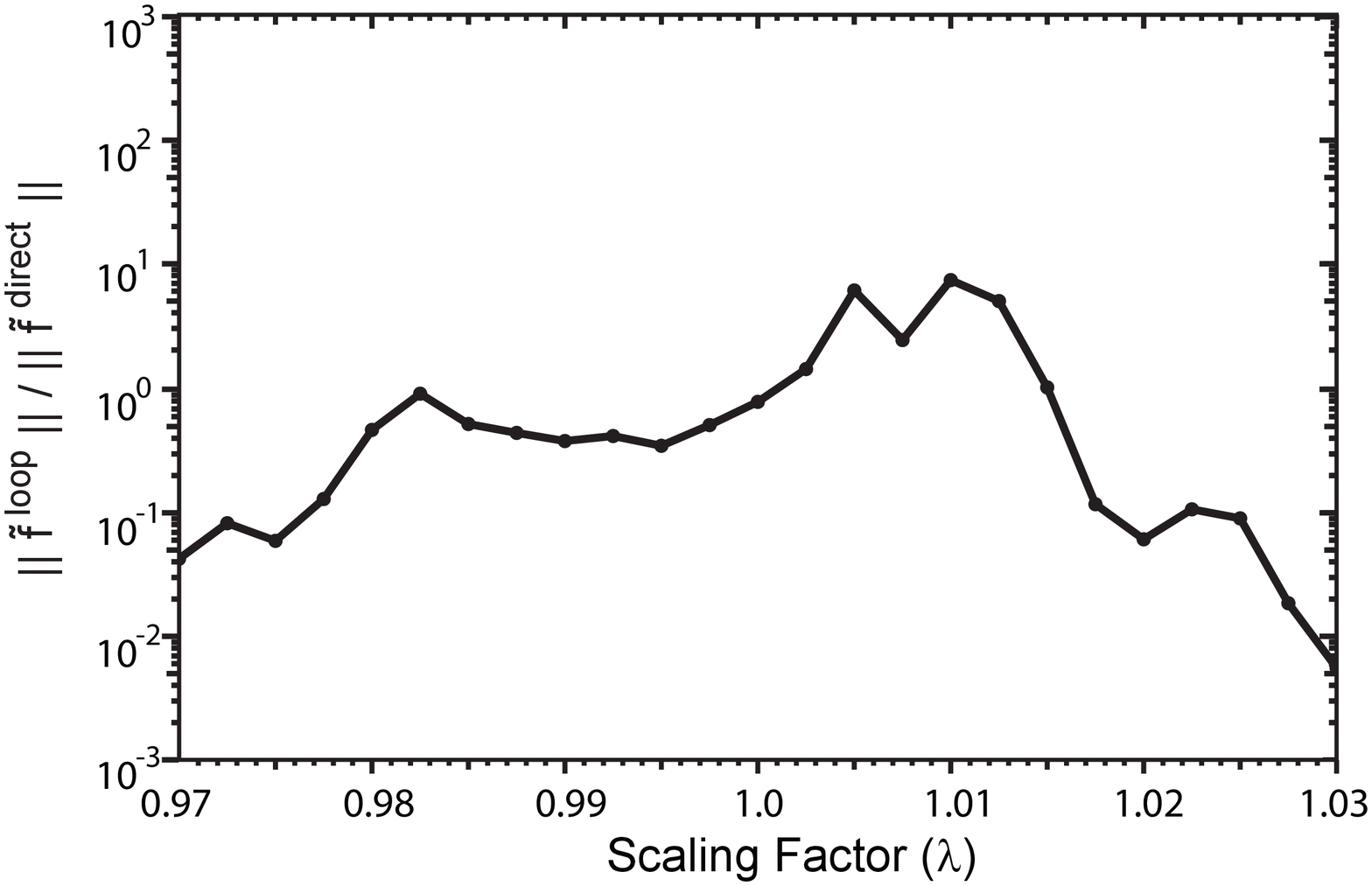}
\caption{
  \begin{small} {\textbf{Three-body potential scaling effects on the magnitudes of the direct and looping scattering amplitudes.}
The ratios of the modulus of the looping (${\tilde f}^{\rm loop}$) and direct (${\tilde f}^{\rm direct}$) scattering amplitudes averaged over the scattering angle are plotted as a function of the 3-body potential scaling parameter at the ultracold collision energy of $1\,{\rm nK}$. The ratios correspond to the rotationally resolved rate coefficients plotted in Fig.~S4.      
    }
\end{small}
}
\label{figS5}
\end{figure}

\begin{figure}[h]
\includegraphics[scale=0.5]{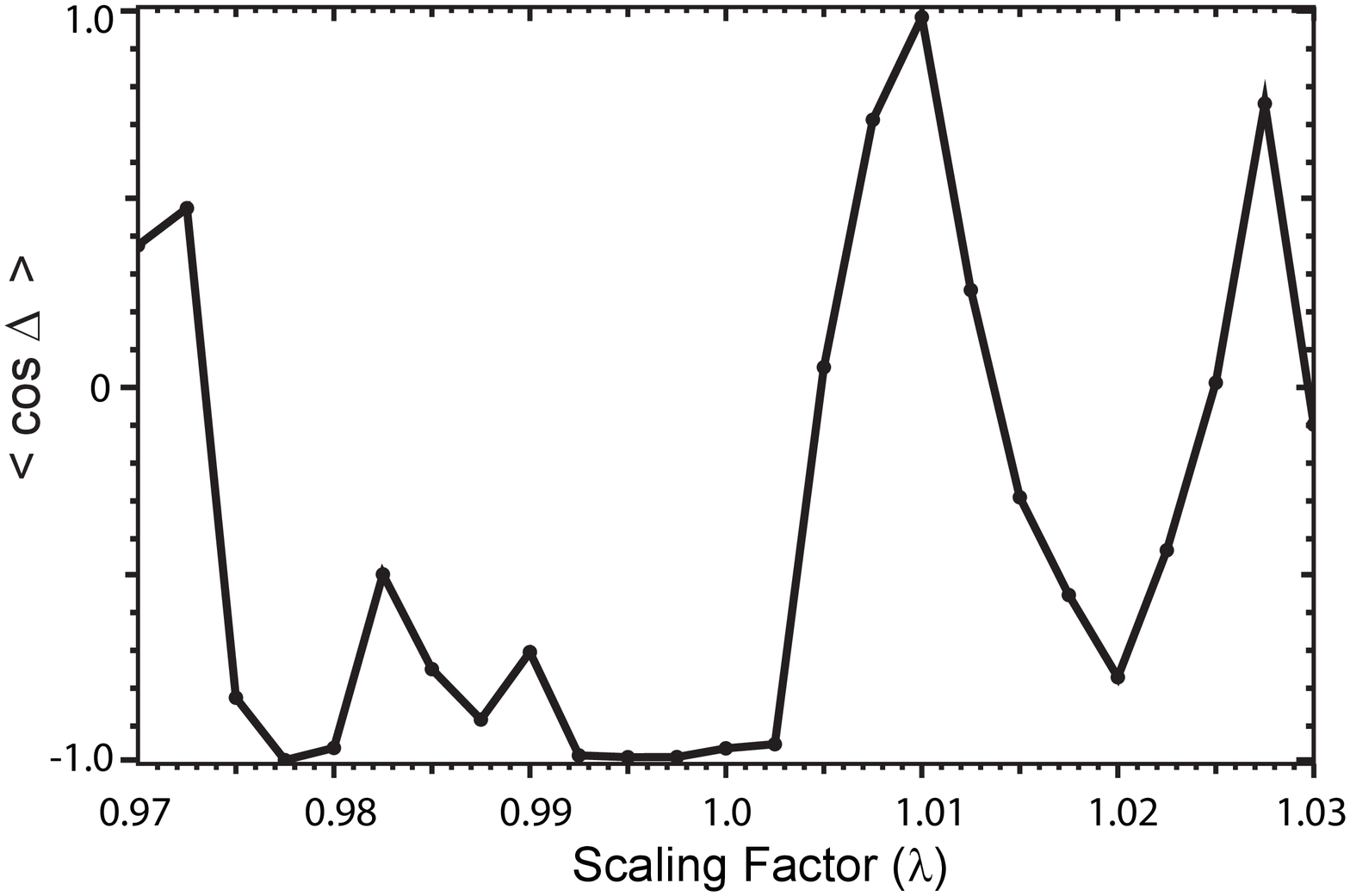}
\caption{
  \begin{small} {\textbf{Three-body potential scaling effects on the relative phase between the direct and looping scattering amplitudes.}
The average $\cos\Delta$ values are plotted as a function of the 3-body potential scaling parameter at the ultracold collision energy of $1\,{\rm nK}$. The $\langle\cos\Delta\rangle$ correspond to the rotationally resolved rate coefficients plotted in Fig.~S4.
      }
\end{small}
}
\label{figS6}
\end{figure}

Armed with Eqs.~\ref{ftotal2_NGPeven} and \ref{ftotal2_NGPodd}, we can now understand and interpret the changes in the rate coefficients plotted in Fig.~S\ref{figS4}.
First consider the scaling range between $\lambda=0.975$ and $1.0025$.
In this region, the $2\times 2$ rate coefficient is significantly enhanced relative to the NGP one (by about $50{\rm x}$ for $\lambda=1$).
This enhancement is due to constructive quantum interference as shown in Eq.~\ref{ftotal2_NGPodd} which implies $n$ must be odd in this region.
Figure S\ref{figS5} plots the ratio of the magnitudes of the looping and direct scattering amplitudes averaged over the scattering angle as a function of the PES scaling factor.
We see that for $0.98 \le \lambda \le 1.0025$ the ratio is near unity which ensures maximal quantum interference.
Figure S\ref{figS6} plots $\cos\Delta$ averaged over the scattering angle as a function of the PES scaling factor.
In the region $0.975 \le \lambda \le 1.0025$ we see that $\langle\cos\Delta\rangle$ lies near $-1.0$ (except for a few values between $0.9825$ and $0.99$).
The negative $\cos\Delta$ is consistent with an odd value of $n$.
As $\lambda$ is decreased below $0.98$, the ratio in Fig.~S\ref{figS5} decreases rapidly to below $10^{-1}$.
Thus, the direct scattering amplitude dominates and quantum interference effects become small.
In this case, the $2\times 2$ and NGP rate coefficients approach each other.
Also in this region, the $\cos\Delta$ reverses sign for $\lambda\le 0.9725$ so that the nature of the quantum interference is reversed and now Eq.~\ref{ftotal2_NGPeven} is relevant (i.e., $n$ becomes even).
Indeed, in this region the NGP rate coefficient is now larger than the $2\times 2$ one.
The change in $n$ from being odd to even (or vice versa) is due to the sudden change in the relative number of bound states between the direct and looping pathways.
Presumably this change is due to a bound (continuum) state leaving (entering) the well and becoming a continuum (bound) state as the well depth is decreased (increased) by the $\lambda$ scaling.
This explains the sudden change in sign on $\cos\Delta$ between $\lambda=0.9725$ and $0.975$.
Now consider the region $1.0025< \lambda \le 1.0125$. In this region the ratios in Fig.~S\ref{figS5} lie between $1$ and $10$ and the $\cos\Delta$ are positive and approach $+1$ for $\lambda=1.01$.
Again the positive $\cos\Delta$ implies an even $n$ (i.e., Eq.~\ref{ftotal2_NGPeven} is relevant) and the NGP rate coefficient is larger than the $2\times 2$ one.
The quantum interference and hence the difference between the rate coefficients is largest in this region for $\lambda=1.01$ where $\cos\Delta\approx 1$.
Between $\lambda=1.0125$ and $1.015$, $\cos\Delta$ changes sign and becomes negative again reversing the nature of the quantum interference (i.e., $n$ becomes odd again).
The $2\times 2$ rate coefficients are now larger than the NGP ones in Fig.~S\ref{figS4}.
As $\lambda$ is increased further, the $\cos\Delta$ oscillates back to being positive again with a peak value at $\lambda=1.0275$.
In the region of large positive $\lambda$, the ratio in Fig.~S\ref{figS5} continues to decrease rapidly which indicates that the direct pathway is dominant.
Thus, the quantum interference decreases considerably and the $2\times 2$ and NGP rate coefficients become comparable in magnitude as seen in Fig.~S\ref{figS4}.
We note that a similar analysis was done for all of the other rotationally resolved rate coefficients and they are all consistent with Eqs.~\ref{ftotal2_NGPeven} and \ref{ftotal2_NGPodd}.

\begin{figure}[h]
\includegraphics[scale=0.5]{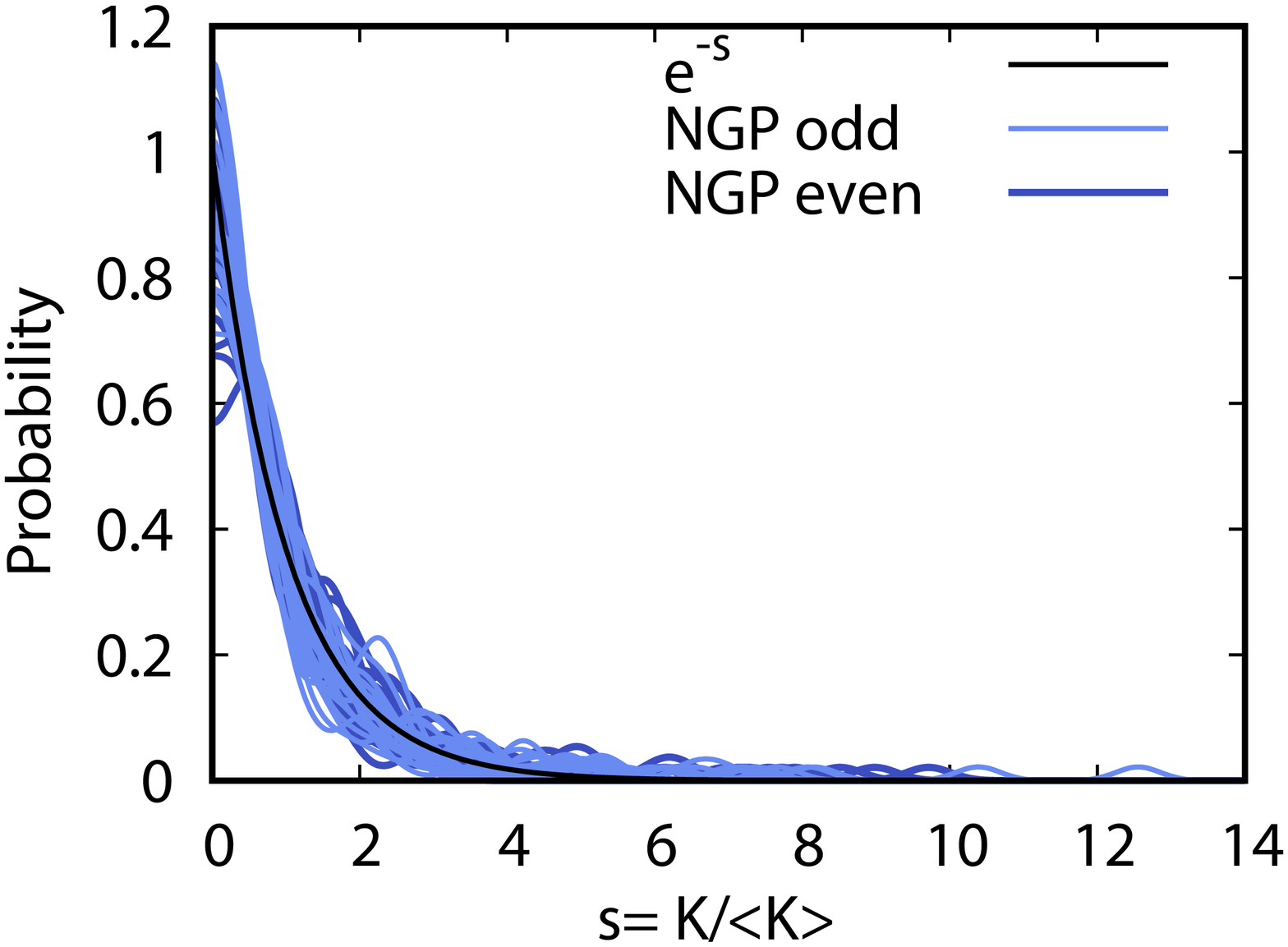}
\caption{
  \begin{small} {\textbf{Probability distributions of NGP rotationally resolved rate coefficients.}
      The probability distributions for all of the rotationally resolved rate coefficients for the Li + LiNa($v=0$, $j=0$) $\to$ Li$_2$($v'$, $j'$) + Na reaction are plotted at the ultracold collision energy of $1.0\,{\rm nK}$. The distributions correspond to the single surface adiabatic (NGP) calculations. The results for even and odd exchange symmetry are plotted in dark and light blue, respectively. Each curve plots the distribution for a different 3-body scaling parameter (see Fig.~S4). The solid black curve is the Poisson distribution.
        }
\end{small}
}
\label{figS7}
\end{figure}

\begin{figure}[h]
\includegraphics[scale=0.5]{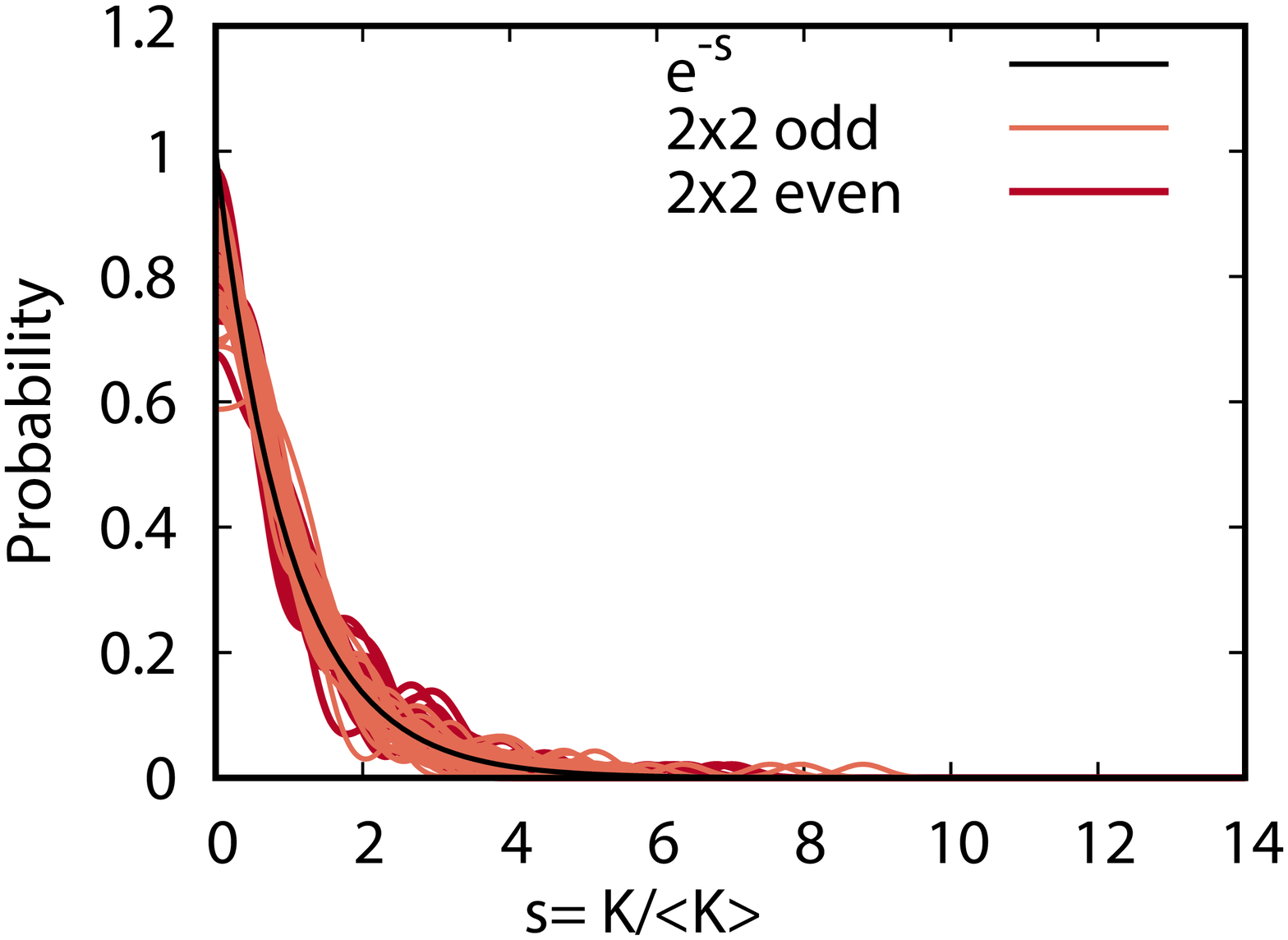}
\caption{
  \begin{small} {\textbf{Probability distributions of diabatic ($2\times 2$) rotationally resolved rate coefficients.}
Same as in Fig.~S7 except that the distributions for the coupled two-state diabatic ($2\times 2$) calculations are plotted. The results for even and odd exchange symmetry are plotted in dark and light red, respectively.
        }
\end{small}
}
\label{figS8}
\end{figure}

Figures S\ref{figS7} and S\ref{figS8} plot the ultracold probability distributions of the rotationally resolved rate coefficients for the
Li + LiNa($v=0$, $j=0$) $\to$ Li$_2$($v'$, $j'$) + Na reaction for the NGP and $2\times 2$ calculations, respectively.
The normalized distributions $s=K/\langle K\rangle$ (where $\langle K\rangle$ denotes the average value of the rate coefficients $K$ for a given data set)
are computed by binning the rate coefficient $K_{v',j'}$ into equally space bins up to 14 times the average value.
In Fig.~S\ref{figS7} (S\ref{figS8}) the even and odd curves are plotted in dark and light blue (red), respectively.
There are $25$ curves for each exchange symmetry in each plot which correspond to the $25$ values of the scaling parameter $\lambda$.
On average, all of the curves are consistent with the Poisson distribution $e^{-s}$ (black curve).

\clearpage
\makeatletter
\renewcommand{\fnum@figure}{\textbf{Fig.~S\thefigure}}
\makeatother

\end{document}